\documentclass[reprint,superscriptaddress,amsmath,amssymb,aps,prx]{revtex4-2}

\usepackage{color}
\usepackage{graphicx}
\usepackage{bm}
\usepackage{physics}
\usepackage{subfig}
\usepackage{xr-hyper}
\usepackage{hyperref}
\usepackage{graphicx}

\newcommand{\TuB}{Institute of Solid State Physics, Technische Universität Berlin, 10623 Berlin, Germany}
\newcommand{\UAM}{Departamento de Física de Materiales, Instituto Nicolás Cabrera, Instituto de Física de la Materia Condensada,
Universidad Autónoma de Madrid, 28049 Madrid, Spain}
\newcommand{\Quandela}{Quandela, 7 Rue Léonard de Vinci, 91300 Massy, France}
\newcommand{\UIBK}{Institute f{\"u}r Experimentalphysik, Universit{\"a}t Innsbruck, Innsbruck, Austria}
\newcommand{\JKU}{Institute of Semiconductor and Solid State Physics, Johannes Kepler University Linz, Linz, Austria}
\newcommand{\Brazil}{Universidade Estadual de Campinas, Instituto de Física Gleb Wataghin, 13083-859 Campinas, Brazil}

\begin{document}

\title{Towards Photon-Number-Encoded High-dimensional Entanglement from a Sequentially Excited Quantum Three-Level System}

\author{Daniel~A.~Vajner}
\affiliation{\TuB}
\author{Nils D. Kewitz}
\affiliation{\TuB}
\author{Martin von Helversen}
\affiliation{\TuB}
\author{Stephen C. Wein}
\affiliation{\Quandela}
\author{Yusuf Karli}
\affiliation{\UIBK}
\author{Florian Kappe}
\affiliation{\UIBK}
\author{Vikas Remesh}
\affiliation{\UIBK}
\author{Saimon F. Covre da Silva}
\affiliation{\JKU}
\affiliation{\Brazil}
\author{Armando Rastelli}
\affiliation{\JKU}
\author{Gregor Weihs}
\affiliation{\UIBK}
\author{Carlos Anton-Solanas}
\affiliation{\UAM}
\author{Tobias Heindel}
\affiliation{\TuB}

\date{\today}

\begin{abstract}
The sequential resonant excitation of a 2-level quantum system results in the emission of a state of light showing time-entanglement encoded in the photon-number-basis -- notions that can be extended to 3-level quantum systems as discussed in a recent proposal \cite{santos2023multipartite}. Here, we report the experimental implementation of a sequential two-photon resonant excitation process of a solid-state 3-level system, constituted by the biexciton-, exciton-, and ground-state of a semiconductor quantum dot. The resulting light state exhibits entanglement in time and energy, encoded in the photon-number basis, which could be used in quantum information applications, e.g., dense information encoding or quantum communication protocols. Performing energy- and time-resolved correlation experiments in combination with extensive theoretical modelling, we are able to partially retrieve the entanglement structure of the generated state.
\end{abstract}

\maketitle

\section{Introduction}
Photonic quantum computers as well as secure quantum networks benefit from the generation of entangled photon states \cite{kimble2008quantum}. Coherently controlled atomic systems have become an excellent source of photonic entanglement over the years \cite{thomas2022efficient}. On the other hand, semiconductor quantum dots (QDs), also known as artificial atoms in the solid-state, provide polarization-entangled photons with high fidelity by exploiting the biexciton (B) -exciton (X) radiative cascade \cite{akopian2006entangled,huber2017highly}. This has already been implemented in experiments for entanglement-based quantum cryptography \cite{basset2019entanglement,basso2021quantum} (see Ref. \onlinecite{vajner2022quantum} for a recent review). Obtaining polarization entangled photon pairs from QDs is, however, not straightforward and requires advanced growth schemes as well as additional external tuning knobs to achieve highest entanglement fidelity \cite{muller2009creating,bennett2010electric,huber2018strain}. So far, using the B-X radiative cascade has been mostly limited to the generation of bipartite entanglement. Exciting the radiative B-X cascade to generate entanglement also imposes intrinsic limitations on the maximum achievable entanglement fidelity and indistinguishability of the emitted photons, due to the finite pulse duration \cite{seidelmann2022two} and the sequential radiative decay \cite{scholl2020crux}, respectively.
\\
Other advanced QD-driving schemes (based on multiple pulses with specific spectral and temporal properties), allow one to obtain entangled photonic states with higher complexity. Exciting a quantum emitter with a sequence of weak laser pulses, i.e. with a pulse area~$\ll\pi$, enables the generation of photonic time-bin entanglement \cite{jayakumar2014time,huber2016coherence,gines2021time}. Moreover, including the polarization degree of freedom of the radiative cascade, photonic hyper-entanglement can be created, which features non-classical correlations between multiple modes thus showing promises to reduce the overhead in some quantum information applications \cite{prilmuller2018hyperentanglement}. Furthermore, the coherent optical driving of dark-exciton- or spin-qubits confined in QDs results in the generation of polarization-encoded cluster-states \cite{schwartz2016deterministic,cogan2023deterministic,tiurev2022high,coste2023high}. These can then be fused to create two-dimensional cluster states \cite{browne2005resource} as resources for fault-tolerant optical quantum computing. Also, the selective driving of dark-exciton states using chirped laser pulses can facilitate the generation of time-bin entangled states \cite{luker2015direct,kappe2024keeping}. 

In the schemes discussed above, the entanglement is entirely created during the QD emission process as a result of the excitonic energy level structure. Alternatively, the entanglement can also be generated externally after the photon emission processes (in the experimental apparatus away from the photon source), e.g., in integrated photonic chips or fiber loops, yielding heralded Greenberger–Horne–Zeilinger (GHZ) states or deterministic linear cluster states, respectively, benefiting from the cavity-enhanced emission of high-performance QD-devices \cite{istrati2020sequential,maring2024versatile,cao2024photonic}.
\\
Among different encoding degrees of freedom to generate quantum states of light, the photon-number basis (or Fock basis) represents a natural and powerful setting to deliver resourceful states for different applications such as quantum communications \cite{arrazola2014quantum}, controlled quantum energy transfer \cite{de2023experimental}, metrology and imaging \cite{munoz2014emitters,munoz2018filtering}, and also perform computationally-hard tasks such as boson sampling \cite{renema2020simulability}. 
In order to encode photonic states in the Fock basis, coherent excitation schemes are required to generate coherent superpositions of different photon number states. While coherent QD excitation schemes, e.g., strictly resonant excitation, produce photon states in such a coherent superposition of different photon-number Fock states \cite{loredo2019generation}, off-resonant or phonon-assisted excitation result in a classical mixture of photon-number states
\cite{bozzio2022enhancing}. Moreover coherence between different photon number states was found to play an important role in quantum information \cite{bozzio2022enhancing}, and can even be controlled by using two-photon resonant excitation (TPE) in combination with an additional stimulation pulse \cite{karli2024controlling}. Applications of photons emitted in a superposition of photon-number states include measurement-device-independent quantum key distribution using photon-number encoding \cite{erkilicc2023surpassing}, or teleportation of photon-number states, as recently demonstrated \cite{polacchi2023quantum}. It has also been reported that coherence between vacuum and 1-photon Fock states has an impact on the evaluation of indistinguishability measurements and on certain quantum circuits \cite{wenniger2024photonic}.
\\
Additionally, photon-number encoding can also be exploited to generate time-entangled states of light, as demonstrated by Wein et al. for the case of a two-level system (2LS) \cite{wein2022photon}. Here, a sequence of two timed, resonant $\pi$-pulses leads to non-classical correlations between the number of photons generated in  the early and the late time-bin mode, yielding a superposition of either two consecutive vacuum states or two consecutive one-photon states. While this scheme can immediately be used to generate maximally entangled bipartite states, it can also be extended to multi-partite states by adding additional excitation $\pi$-pulses to the sequence, causing an increase in the number of temporal modes with a structure following the Fibonacci sequence \cite{wein2022photon}.
\\
In this work, we experimentally demonstrate a path to increase the entanglement-dimension of a photon-number-encoded state, not by adding more excitation pulses, but by using a solid-state 3-level system (3LS) and adding energy as a degree of freedom, following a recent theory proposal by Santos et al. \cite{santos2023multipartite}. In the following, we briefly review the implemented excitation protocol, before providing both analytical as well as numerical predictions for the behavior of the emitted entangled state. Finally, we present a detailed experimental study analyzing the mode-structure of the generated photonic state.

\section{Entanglement Generation Protocol}
\begin{figure*}[ht]
    \centering
\includegraphics[width=1\linewidth]{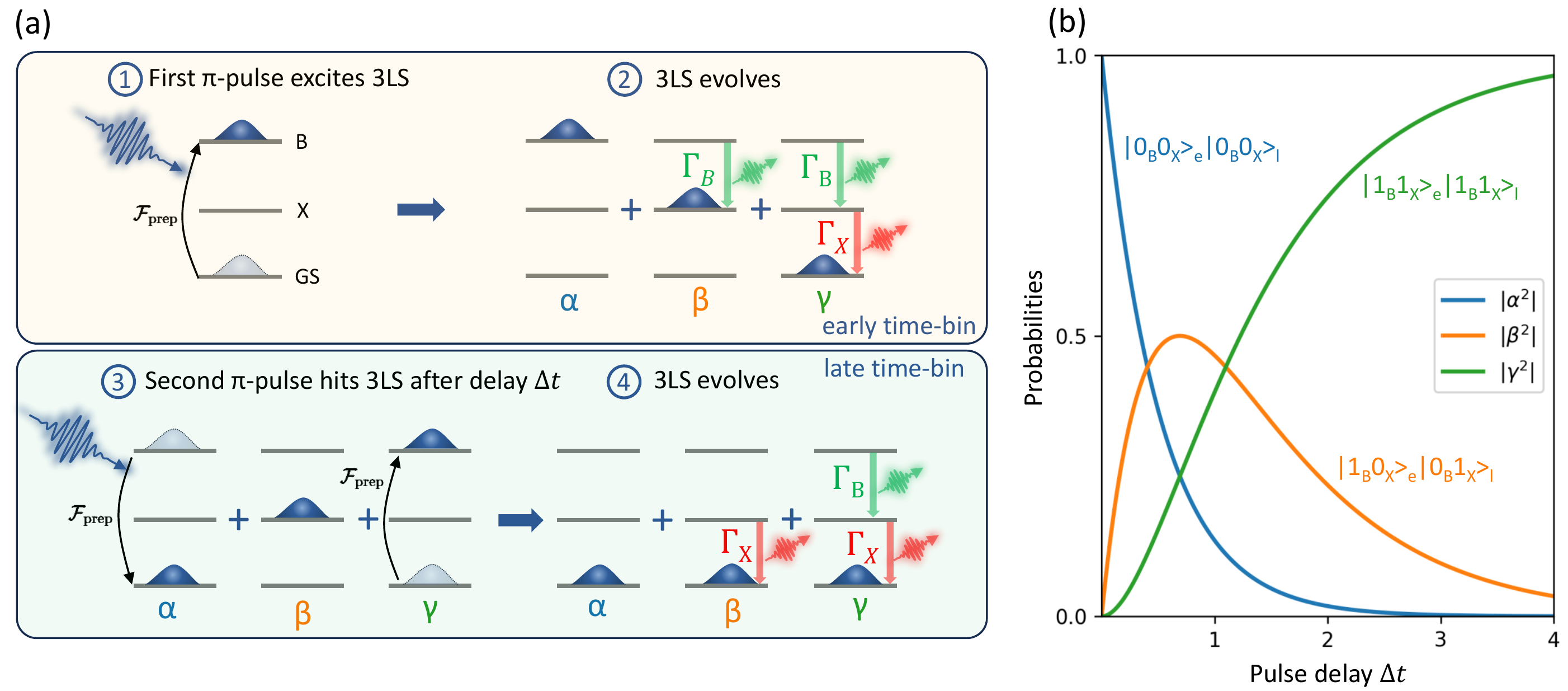}
\caption{(a) Schematic of the sequential TPE protocol: \textbf{1.} Initial TPE via $\pi$-pulse brings the system to the B state with probability $\mathcal{F}_{\text{prep}}$. \textbf{2.} The system evolves freely until the second pulse arrives after delay $\Delta t$ during which it can stay in B state, decay to X state emitting a photon or decay fully to GS emitting 2 photons in the early time-bin. \textbf{3.} The second TPE $\pi$-pulse arrives and depending on the system state creates a superpositon of de-excitation, unchanged X state or re-excited B state. \textbf{4.} Final decay leads to emission of zero, one or two photons in second time bin. Occupied (empty) states are marked in dark (light) blue. (b) Populations for $\Gamma_X=2\Gamma_{B}$ as a function of the delay between the TPE pulses for the ideal case of $\mathcal{F}_{\text{prep}}=1$, indicating full de-excitation for small $\Delta t$ and double excitation for large $\Delta t$. The labels indicate the contribution of the respective terms to the state $\ket{\psi}$.}
\label{fig:schematic}
\end{figure*}

In this work the quantum 3LS is realized by a single semiconductor QD excited via TPE \cite{jayakumar2013deterministic,muller2014demand,huber2017highly}, resulting in a coherent population of the excited B state with near unity fidelity under $\pi$-pulse excitation. After a temporal delay $\Delta t$, a second $\pi$-pulse with the same color, i.e., two-photon resonant with the ground state (GS) to B transition, is launched onto the QD. In this situation, the effect of the second laser pulse on the 3LS will depend on the state of the QD. Thus, we can distinguish the following three distinct cases when the second pulse arrives (see schematic in Figure~\ref{fig:schematic}(a)):
\begin{itemize}
    \item[(i)] The 3LS is still excited in the B state: The second $\pi$-pulse stimulates the complete de-excitation, resulting in the emission of a vacuum state with probability~$|\alpha^2|$.
    \item[(ii)] The 3LS has decayed to the X state: The second pulse is energy detuned and has no effect, a B photon in the first time-bin and an X photon in the second time-bin are emitted with probability $|\beta^2|$.
    \item[(iii)] The 3LS has fully decayed to the GS: The second pulse re-excites the B state, resulting in the emission of two B-X photon pairs with probability $|\gamma^2|$.
\end{itemize}
The final state thus is a superposition of the above described three cases weighted with their respective delay-dependent probabilities $|\alpha(\Delta t)|^2, |\beta(\Delta t)|^2, |\gamma(\Delta t)|^2$. By partitioning the time into two modes, before (early time-bin: e) and after (late time-bin: l) the arrival of the second laser pulse, the final photonic state reads as follows \cite{santos2023multipartite}:
\begin{eqnarray}
    \ket{\psi} &=& 
    \underbrace{\alpha\left|0_B 0_X\right\rangle_e\left|0_B 0_X\right\rangle_l}_{(i)}+\underbrace{\beta\left|1_B 0_X\right\rangle_e\left|0_B 1_X\right\rangle_l}_{(ii)} \nonumber \\
    &+& \underbrace{\gamma\left|1_B 1_X\right\rangle_e\left|1_B 1_X\right\rangle_l}_{(iii)} \,\,.
    \label{eq:state}
\end{eqnarray}
 Eq.~\ref{eq:state} describes an energy- and time-entangled state, encoded in the photon-number basis, which lives in a 16-dimensional Hilbert space. It is composed of 4 modes, two energy modes (B,X) and two time modes (e,l), each representing a two-level system composed of the vacuum $ \ket{0}$ and the single photon $\ket{1}$ state. The multi-partite entanglement of the state $\ket{\psi}$ can be quantified by computing the mutual information between different partitions of the state as described in Ref. \onlinecite{santos2023multipartite}. Given two reduced density matrices $\rho_1$ and $\rho_2$ that correspond to two sub-spaces of the joint state $\rho_{1,2} = \ket{\psi} \bra{\psi}$, and the von-Neumann entropy $S(\rho)=-\text{tr}[\rho \, \text{log}_2(\rho)]$, one can calculate the mutual information between the two parties as $I(\rho_1:\rho_2) = S(\rho_1)+S(\rho_2)-S(\rho_{1,2})$. By evaluating this for all the different possible partitions one finds that some partitions yield a mutual information higher than that of the GHZ state, as shown by Santos et al. \cite{santos2023multipartite}.
 \\
 Note, that by including the polarization as an additional degree of freedom, the dimension of the state will be further increased to 32 allowing for more complex partitions. In the experiments conducted in this work, however, we only collect one polarization state for the sake of simplicity. Moreover, our protocol may be further extended using additional delayed $\pi$-pulses, adding more multi-partitions to the temporal bins, hence increasing the complexity of the entangled state. 

The relative weights of the state $\ket{\psi}$ follow from the occupation probabilities during the B-X cascade. In the ideal case without additional dephasing mechanisms or non-radiative decay channels influencing the cascade, the probability $|\alpha|^2$ to stay in the B state decays mono-exponentially. Then, the X state is only shortly occupied with probability $|\beta|^2$ before it decays, while the probability $|\gamma|^2$ captures the occupation of the ground state (see dynamics in Figure~\ref{fig:schematic}(b)). With the radiative decay rates $\Gamma_{B,X}=1/\tau_{B,X}$, for the B and X states respectively ($\tau_{B,X}$ are the corresponding lifetimes), one obtains the probabilities for the three terms in the $\ket{\psi}$ state as:
\begin{eqnarray*}
    |\alpha (\Delta t)|^2&=&e^{- \Gamma_B \Delta t} \\
    |\beta (\Delta t)|^2 &=&\frac{\Gamma_B\left(e^{-\Gamma_B \Delta t}-e^{-\Gamma_X \Delta t}\right)}{\Gamma_X-\Gamma_B} \\
    |\gamma (\Delta t)|^2 &=& 1-\alpha^2-\beta^2 \,\,.
\end{eqnarray*}

Note that for semiconductor QDs the biexciton state typically decays faster than the exciton state ($\Gamma_B>\Gamma_X$), constraining the possible values of the terms in $\ket{\psi}$. The well known GHZ state $(\ket{0000}+\ket{1111})/\sqrt{2}$, for instance, is not easily accessible, as it would require a vanishing $\beta$ which implies a much faster X- than B-decay, which either requires asymmetric Purcell enhancements via engineered photonic environments \cite{bauch2023demand} or very specific quantum dot properties \cite{reischle2008influence}.

\section{Results}
\subsection{Analytical and Simulated Predictions}
In our experiments the emitted multi-partite state $\ket{\psi}$ will be probed by measuring the photon arrival time distribution, also allowing the retrieval of the emitted average photon number $\mu_{B,X}(\Delta t)$, and the second-order correlation function $g_{i,n-j,m}^{(2)}(\Delta t)$, where the sub-indices $i,j = \{B,X\}$ ($n,m= \{e,l \}$) correlate two energy- (time)- modes, respectively. Moreover, the wave-packet interference in Hong-Ou-Mandel (HOM)-type experiments is needed to confirm the purity of the state $\ket{\psi}$ as only the phase coherence between the vacuum and single photon components enables entanglement between different partitions of the state \cite{santos2023multipartite}, a point to be addressed later in this work. For the sake of simplicity, in the following, we will consider these quantities to be resulting from instantaneous pulses that perfectly control the biexciton level, in a 3LS -- free from external sources of dephasing, e.g., due to charge-, phonon-, or spin-noise.

The mean photon number $\mu$ can be described in the different energy and time modes or integrated over e.g. the temporal modes, resulting in the mean number of photons per energy mode. This average photon number $\mu_{B,X}(\Delta t)$ of the different energy modes will increase with increasing time delay $\Delta t$ between the two laser pulses, from (ideally) 0 emitted photons for $\Delta t = 0$, in the case of complete de-excitation of the biexciton (Eq.~\ref{eq:state}, term $(i)$), to 2 emitted photons per energy mode for $\Delta t \gg \tau_X $, in the limit of fully separate, sequential excitation events (Eq.~\ref{eq:state}, term $(iii)$). The expected mean photon number can be calculated as $\mu_i= \bra{\psi}  a_i^\dagger a_i \ket{ \psi}$ ($i=\{B,X\}$) yielding $\mu_{B,X}(\Delta t)= \beta^2(\Delta t) + 2 \gamma^2(\Delta t)$ for the mean photon number in each energy mode. 
Because the energy modes are easier to separate experimentally, we will compare this expression to the measured number of photons in each energy mode later (cf. Figure~\ref{fig:lifetimes}).
But note that the same expression also describes the mean photon number of photons in each temporal mode $\mu_{e,l}$ due to the symmetry of $\ket{\psi}$.\\

The population probabilities $|\alpha|^2$, $|\beta|^2$, and $|\gamma|^2$ in Eq.~\ref{eq:state} can be retrieved via time- and energy-resolved cross-correlation experiments. For the ideal state $\ket{\psi}$, these correlations can be calculated analytically, via the expected second-order correlation function
\begin{equation}
    g_{i,n-j,m}^{(2)}(\Delta t) = \frac{\bra{\psi} a^\dagger_{i,n}a^\dagger_{j,m}a_{j,m}a_{i,n}\ket{\psi}}{\bra{\psi}a^\dagger_{i,n}a_{i,n}\ket{\psi} \bra{\psi}a^\dagger_{j,m}a_{j,m}\ket{\psi}} \,\,.
    \label{eq:g2}
\end{equation}
Evaluating Eq.~\ref{eq:g2} for all possible combinations of the energy- and time-modes $i=\{B,X\}$ and $j=\{e,l\}$ yields that the normalized coincidence probability of finding two photons in the same mode $g^{(2)}_{i,n-i,n}(\Delta t) = 0$, as expected for the single-photon emission from biexciton to exciton or exciton to ground state in a given time-bin. Correlations describing the coincident detection of a biexciton in the late and an exciton in the early time-bin can only be caused by re-excitation of the 3LS (Eq.~\ref{eq:state}, term ($iii$)) and are hence described by $g^{(2)}_{B,l-X,e}(\Delta t)=1/\gamma^2$. All other correlations behave as $1/(\beta^2 + \gamma^2)$ as they can be caused by both terms ($ii$) and ($iii$) in Eq.~\ref{eq:state}. The experimentally measured data will later be compared to these analytical predictions, obtained under the ideal conditions described in the beginning of this section. 

In practice, however, our experiments face certain imperfections, related to both the QD physics as well as the measurement apparatus itself, which cannot be included directly into the analytical calculations. Firstly, the QD is not an ideal 3LS. It is subject to different dephasing mechanisms (as noted above) and potentially subject to an imperfect coherent control of the biexciton with the $\pi$-pulse laser. This was evidenced by measuring a finite preparation fidelity ${\mathcal{F}_{\text{prep}} \approx 0.87}$, obtained from B-X cross-correlation measurements \cite{wang2019demand,neuwirth2022multipair,vajner2024demand}. In combination with phonon-interactions, this will limit the amount of de-excitation caused by the second TPE pulse. Secondly, our experimental setup adds imperfections such as: (1) the detector timing jitter $\delta t \approx 40\,\text{ps}$,  (2) the detector dead-time  ${t_{\text{Dt}}=100\,\text{ns}}$ and (3) the finite temporal length of the laser pulse $\tau_{\text{Laser}} \approx 6\,\text{ps}$. Additional experimental imperfections to consider are the polarization-filtering in the detection path (we detect only one polarization, which enhances correlations within one cascade), slightly unbalanced beamsplitter (BS) ratios $r_{\text{BS}}$, and setup transmission losses. 
\\
To enable a comparison of our experimental data with theoretical expectations, we therefore implement stochastic Monte Carlo (MC) simulations accounting for all above-mentioned experimental imperfections including a finite preparation fidelity. The simulations are based on the assumptions that all radiative decays follow an exponential probability distribution with the decay times being determined from the experiment, and that the single photon emitted from the exciton is always emitted after the biexciton photon. In each simulated excitation cycle, an excitation or de-excitation event occurs with a probability that equals the preparation fidelity $\mathcal{F}_{\text{prep}}$. Next, for excited states the simulation randomly draws a value for the decay time (from the exponential distribution), a random polarization (maintaining the correlation for photons within the same cascade), and randomly chooses on which detection channel the event will be registered. The timing jitter is incorporated by randomly drawing a number from a Gaussian distribution of width $\delta t$, which is then added to the arrival time at the respective detection  channel. We note that such simulations do not describe the individual dynamics on a quantum level. Nevertheless, they allow us to not only reproduce the same conditions as in the experiment, but also to vary different experimental parameters to observe their effect on the expected correlations, which facilitates a detailed understanding of limiting factors for the entanglement generation scheme employed here. Specifically, it enables us to simulate the two-time-correlation maps that one would expect for infinitely fast detectors, a perfect preparation (and depopulation) fidelity, and a lossless setup without polarization filtering, and use them to validate our data evaluation workflow.
\\
While the photonic state populations in Eq.~\ref{eq:state}, accessible via mode-dependent two-photon correlations (see Eq.~\ref{eq:g2}), correspond to the diagonal elements of the state's density matrix, it is necessary to also assess the off-diagonal elements of the state to quantify its purity, as also a mixed state may exhibit the same diagonal elements. 
\\
Note that in principle a full tomography of the state $\ket{\psi}$ is needed: But as that includes the measurement of 4-photon interference and 4-fold coincidence detections it is experimentally not feasible in this work. Therefore, we will argue in the following why the purity of the state $\ket{\psi}$ is expected and discuss how the coherence between zero and one photon states can be observed in phase-resolved interference experiments.
\\
Since the observation of Rabi rotations between the biexciton- and ground-state confirms the coherent superposition between the electronic states, the TPE scheme also yields pure photonic states after the spontaneous emission. For example, a single TPE pulse with a pulse area $\theta$ generates an energy-entangled state encoded in the photon-number basis of the form $\cos{(\theta/2)}\left|0_B 0_X\right\rangle+\sin{(\theta/2)} \left|1_B 1_X\right\rangle$. This coherent driving has enabled the generation of time-entangled states $1/\sqrt{2}(\left|1_B 1_X\right\rangle_e+e^{i\phi}\left|1_B 1_X\right\rangle_l)$ in previous works (with $\theta \ll \pi$) \cite{jayakumar2014time,huber2016coherence,gines2021time}. 
This, however, does not imply that the spectrally filtered state of only the X or B emission is a pure state in the photon-number basis since the individual energy modes are maximally mixed, as confirmed in experiments \cite{karli2024controlling}. Nevertheless, the initial coherence present between B, X, and vacuum after the one-pulse TPE excitation is a necessary condition to obtain the pure state $\ket{\psi}$ from the sequential excitation.
\\
A way to partially characterize the purity of the full state $\ket{\psi}$ is the implementation of a phase-controlled HOM interference of two copies of the state. This HOM interference unveils coherence between vacuum and single-photon states by interfering subsequently emitted versions of $\ket{\psi}$ in a path-unbalanced Mach-Zehnder interferometer (MZI) (cf. inset of Figure \ref{fig:HOM}(a)), where the relative phase $\phi$ between the two arms is controlled \cite{wein2022photon}. In the following, we analytically predict the expected behavior.
\\
The HOM experiment will yield correlations of the form $g^{(2)}_{\text{HOM,i,n--j,m},}(\Delta t,\phi)$ that depend on the relative phase~$\phi$. To calculate the theoretical expectations of interfering two ideal $\ket{\psi}$ states at a lossless BS, a toy model is used assuming that the BS inputs are described by the tensor product $\ket{\psi_{\text{in}}}= \ket{\psi}_a \otimes \ket{\psi(\phi) }_b$, where \textit{a} and \textit{b} are the two spatial input modes of the BS. Here, the relative phase $\phi$ is picked up by each photon in mode \textit{b}, due to the different optical path length with respect to mode \textit{a} in the MZI before the interference at the BS. After applying the corresponding BS transformations, the output state $\ket{\psi}_{\text{HOM}}$ contains up to eight photons, distributed over the two available spatial $\{c,d\}$, spectral \{B,X\}, and temporal $\{e,l\}$ modes. Note that this is a simplification neglecting the action of the first BS in the HOM interferometer. As long as temporal post-selection restricts the evaluation to the zero-delay-peak this is valid. But when normalizing with respect to the side peaks one would need corrections as discussed recently for superpositions of photon number states in Ref. \cite{wenniger2024photonic}.
\\
Again, the expected inter-mode correlations can be calculated in analogy to Eq.~\ref{eq:g2}. Here, the expected correlations after interfering two copies of $\ket{\psi}$ originate from an interplay of (1) the HOM bunching of indistinguishable photons in the same energy and time mode, and (2) the phase-dependent interference between different superpositions of zero and one-photon components. This leads to interesting consequences, such as the fact that while only photons in the same energy and time modes interfere, manifestations of the phase-dependent interference are most strongly observed in correlations between different modes, as also reported in Ref. \onlinecite{wein2022photon}.
\\

In practice it is challenging to interfere two subsequently emitted $\ket{\psi}$ states, while fully resolving the correlation between the different spatial, temporal, and spectral modes. Interestingly, resolving all modes in the HOM detection is not necessary to, at least, witness an influence of the relative phase $\phi$ and thus to exclude the presence of a fully mixed state in the Fock basis. Even for the simplest case, when neither the spectral nor the temporal modes are resolved and only correlations between the BS outputs are measured (like in a standard HOM measurement), one already expects phase-dependent correlations of the form
\begin{eqnarray}
g^{(2)}_{\text{HOM}}(\Delta t) &=& \frac{2\beta^4+13\beta^2 \gamma^2 + 12\gamma^4+\alpha^2(\beta^2+6\gamma^2)}{4(\beta^2+2\gamma^2)^2}\nonumber \\
&-& \frac{\beta^2(\alpha^2+\gamma^2)\cos(2\phi)}{4(\beta^2+2\gamma^2)^2}\,\,.
    \label{eq:hom}
\end{eqnarray}
We note again that this analytical calculation assumes perfect indistinguishability, while it is well known that the cascaded radiative decay limits the indistinguishability of the emitted B and X photons \cite{scholl2020crux}, reducing the oscillation amplitude of $g^{(2)}_{\text{HOM}}(\Delta t,\phi)$ as a function of $\phi$. 
Thus, to predict the results of the phase-dependent interference of a realistic $\ket{\psi}$ state, we also perform additional numerical simulations  by solving the open system master equation for the realistic 3LS excited by a sequence of instantaneous $\delta$-pulses. In this way, we include the timing jitter from the cascaded decay reducing the indistinguishability. We find that these simulated correlations qualitatively agree with the analytical prediction of the ideal case, while they quantitatively deviate due to the reduced indistinguishability.

\subsection{Measuring Lifetime and Intensity-Correlations}
In our experiments, the quantum 3LS is realized by a single GaAs/AlGaAs semiconductor QD, excited with a sequence of two delayed TPE pulses (see Methods Section for details on the sample fabrication and characterization). The coherent nature of the state preparation is thereby confirmed by the observation of Rabi rotations in the B and X emission and we determine a  preparation fidelity of $F_{\text{prep}} \approx 87\%$ (via B-X cross-correlation experiments under single-pulse TPE \cite{vajner2024demand}). The QD used in the following shows radiative decay times of $\tau_B = 142(1) \, \text{ps}$ and $\tau_X=187(1) \, \text{ps}$ as well as a B-X binding energy of 3.59(1)$\,$meV. Single-photon emission is confirmed when exciting with a single TPE $\pi$-pulse by anti-bunching values of $g_B^{(2)}(0) = 0.02(1)$ and $g_X^{(2)}(0) = 0.03(1)$ for the B- and X-state, respectively, limited by scattering of residual laser light. Moreover, we measure a photon-indistinguishability of 42(2)$\,\%$ in HOM experiments with X photons separated by a 12.5$\,$ns delay. The indistinguishability is close to the limit of $V_{\text{max}} = 1/(1+\tau_B/\tau_X) \approx 57\%$ expected from the measured radiative lifetime ratio \cite{scholl2020crux}. The indistinguishability of the B photon is typically lower bounded by the result of the X measurement \cite{muller2014demand,huber2017highly,wang2019demand,liu2019solid,scholl2020crux}.

\begin{figure*}[ht]
\centering
\includegraphics[width=1\linewidth]{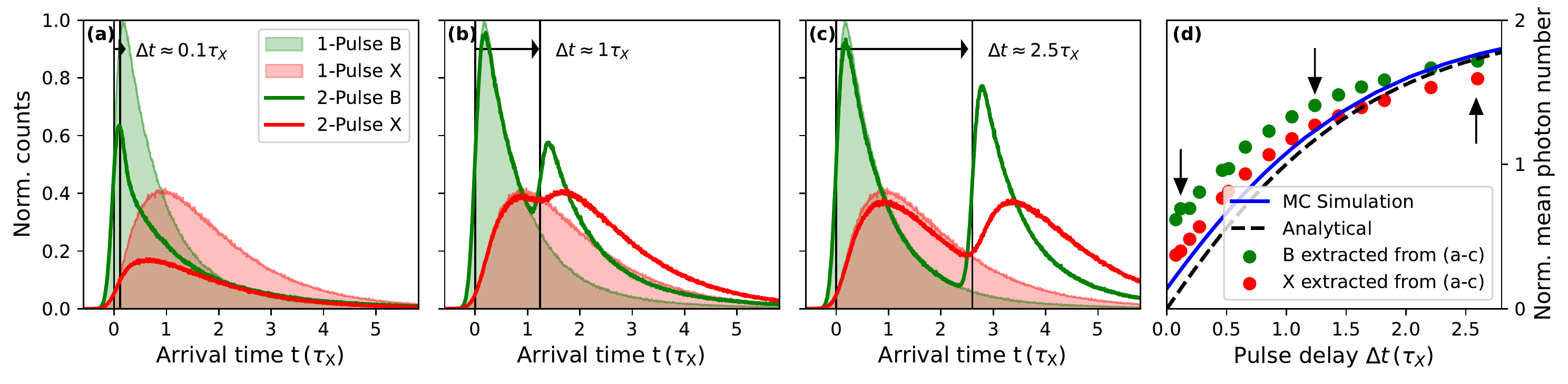}
\caption{(a-c) Experimentally measured photon arrival time distributions under sequential TPE (pulse duration 6$\,$ps corresponding to $\tau_X/31$) for different excitation pulse separations $\Delta t$ showing the B photons (green), the X photons (red) as well as the corresponding measurement with only one excitation pulse for comparison (grey). The de-excitation ($\Delta t \ll \tau_X$) and re-excitation ($\Delta t > \tau_X$) can be observed. (d) Integrating over all events for each energy mode yields the measured mean photon numbers, normalized by their respective one-pulse results. The data is compared to the analytical expression of $\mu_{X,B}(\Delta t)$ (black dashed line) as well as the MC simulated events (blue solid line). Black arrows indicate the three different $\Delta t$ values in (a-c).}
\label{fig:lifetimes}
\end{figure*}

To confirm the generation of the $\ket{\psi}$ state under two-pulse TPE driving, we first studied the decay dynamics of the cascade followed by the two-time correlation histograms via generalized Hanbury-Brown \& Twiss type experiments.\cite{brown1956correlation} Figure~\ref{fig:lifetimes} (a-c)) shows the time-traces of the X (red line) and B (green line) emission for different temporal delays $\Delta t$ between two sequential TPE pulses (marked by black vertical lines). The transparent curve corresponds to the single TPE pulse time-trace. The experimental data confirm the resulting de-excitation effect of the second pulse on the B-state as well as the reduction of the X emission. For $\Delta t \ll \tau_X$ we observe a clear B de-excitation, resulting in a reduction of the average number of emitted photons, compared to the case with only one excitation pulse (with $\mu \approx 1$, see filled curves). For $\Delta t \gg \tau_X$ we observe two quasi-uncorrelated excitation events.
\\

Figure~\ref{fig:lifetimes}(d) presents a quantitative analysis of the mean photon number $\mu_{X,B}(\Delta t)$ in each spectral mode as a function of $\Delta t$ (see green and red data points, respectively); in these datasets the integrated emission of the two-pulse excitation is normalised to the single-pulse emission, and compared to theory predictions. The experiments clearly follow the simulations, showing a suppression of $\mu_{B,X}$ at small $\Delta t$, and the convergence to an average photon number of two for large $\Delta t$. The experimental deviation from the analytical expression for $\mu_{X,B}(\Delta t)$ at small $\Delta t$ is attributed to the imperfect preparation fidelity, as the minimum mean photon number can be described by $\mu_{X,B}(0) = 1- \mathcal{F}_{\text{prep}}$ if other imperfections can be neglected.  Additionally, the finite pulse duration of the excitation pulse, i.e., $0.03\, \tau_X$ in our experiments, imposes a lower limit on $\Delta t$ to avoid overlap (and interference) of the two excitation pulses. The slight deviation at high $\Delta t$ might be caused by small variations of the second pulse power when changing $\Delta t$.

\begin{figure*}[ht]
    \centering \subfloat{\includegraphics[width=0.9\linewidth]{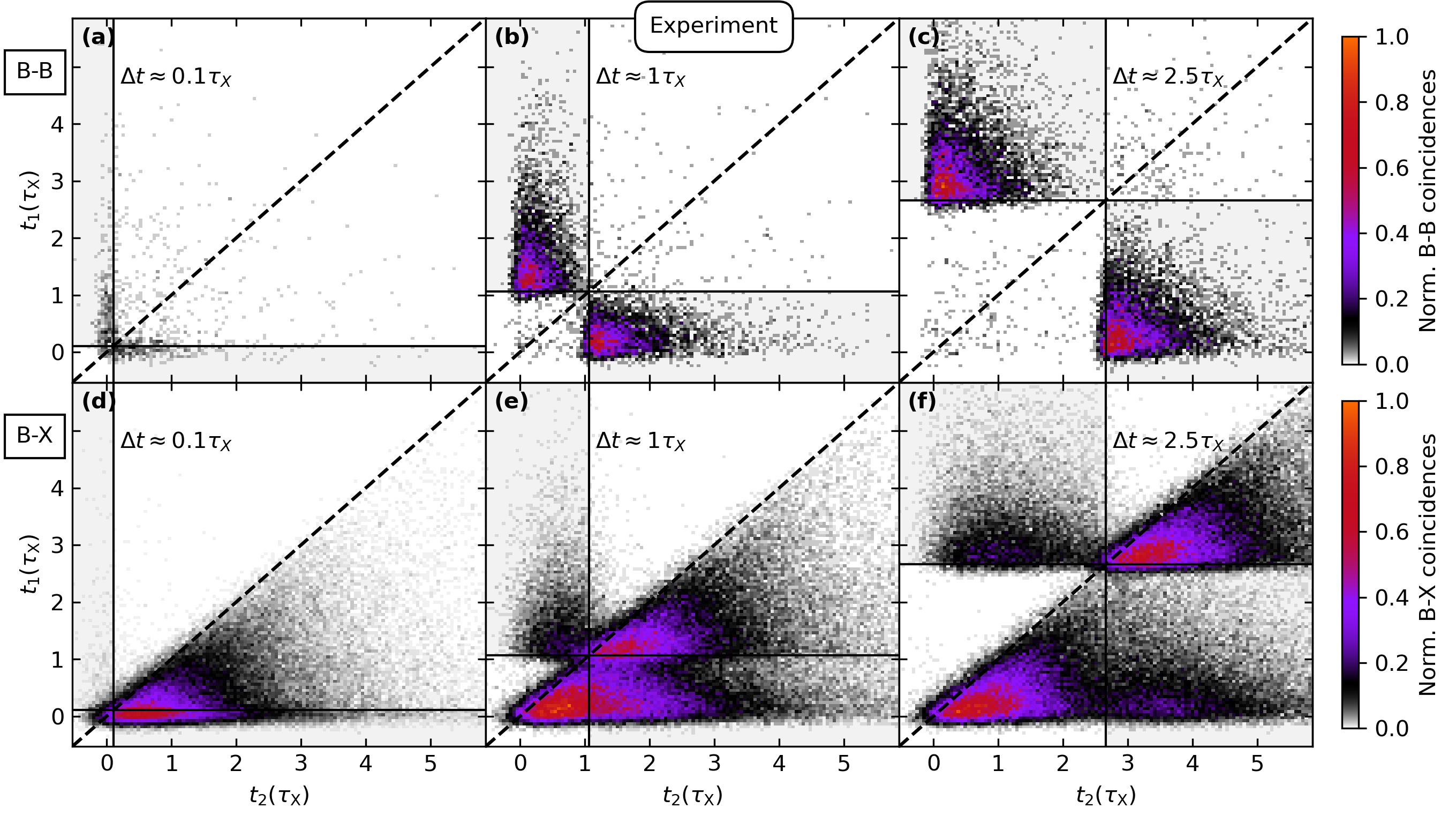}}\\\subfloat{\includegraphics[width=0.9\linewidth]{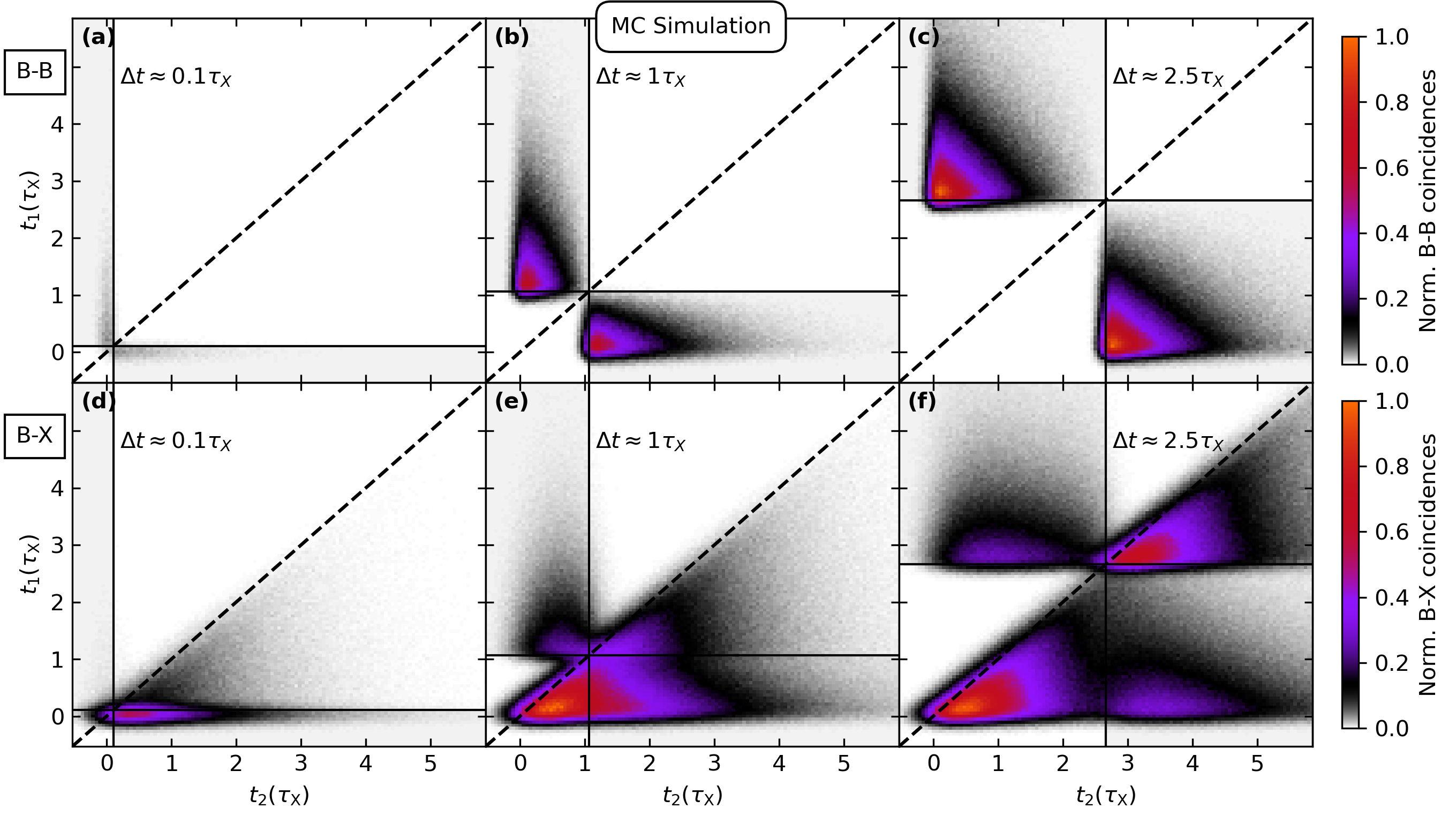}}\\
    \caption{Top: Experimentally measured two-time correlation histograms within the same energy mode (a-c) and between different energy modes (d-f),for different pulse separations $\Delta t$, show distinct features of the sequential TPE excitation. The black line separates the different time bins defining four quadrants corresponding to the time-resolved coincidence cases: $\{e-e, e-l, l-e, l-l\}$. Bottom: Monte Carlo (MC) simulation including realistic experimental parameters qualitatively reproduces the features of the experiment.}
    \label{fig:butterflies_data}
\end{figure*}

Now that the coherent control via the second delayed laser pulse is confirmed, the next steps concern the exploration of the time ordering as well as the correlations among the generated photons in the different modes. Using two-time correlation measurements, we probe the correspondence of the different photons of the $\ket{\psi}$ state to their respective temporal and spectral modes.\cite{wein2022photon} The $\ket{\psi}$-state is analyzed by correlating the arrival times of detected photons on two detectors ($t_1$ and $t_2$) for the case of photons from the same (B-B or X-X) as well as for different (B-X or X-B) spectral modes and the results are visualized in two-time correlation maps. Figure~\ref{fig:butterflies_data} shows the two-time correlation maps for the B-B and B-X mode-combinations for the experimental data (top panels) and the simulated data from the Monte Carlo simulation (bottom panels). The $\Delta t$ values correspond to the delay values depicted in Figure~\ref{fig:lifetimes}. The time-bin assignment is visualized by horizontal and vertical black lines, dividing each two-time correlation map into four quadrants $\{ee,el,ll,le\}$. In the following we comparatively discuss the experimental and simulation data.
\\
Several distinct features of the sequential excitation can be noticed. For correlations within the same energy mode (Subfigures~\ref{fig:butterflies_data}(a-c) for experiments and simulations, respectively) one expects no coincidences within the same time bin (\textit{e-e} and \textit{l-l}) due to the single-photon nature of the emitter. Instead, coincidences between single photons from different time bins (in the \textit{e-l} and \textit{l-e} quadrants) are possible due to the 3LS re-excitation. The correlations between different energy modes (Subfigures~\ref{fig:butterflies_data}(d-f) for experiment and simulation, respectively) show a strong asymmetry caused by the sequential B-X emission; for $\Delta t \ll \tau_X$ (cf. Subfigures~\ref{fig:butterflies_data}(d)) no X photon can arrive before the B photon, explaining the concentration of events only below the diagonal dashed line. Additionally, the fact that $\tau_X > \tau_B$ creates an additional asymmetry of the correlation shape within all quadrants of the B-X correlations. For $\Delta t \geq \tau_X$ (cf. Figure~\ref{fig:butterflies_data}(e)) also events above the diagonal line occur, indicating coincidences between an early X and a late B. This happens whenever a B photon of the first cascade is not detected, only the X, followed by the detection of a B photon from the re-excited second cascade, corresponding to the (iii) term in Eq.~\ref{eq:state}. We note that all these experimental features are well reproduced in our MC simulations displayed below the experimental results; these simulations are conducted using our experimental parameters (see Methods for more details).

\begin{figure*}[ht]
    \centering
\subfloat{\includegraphics[width=0.99\linewidth]{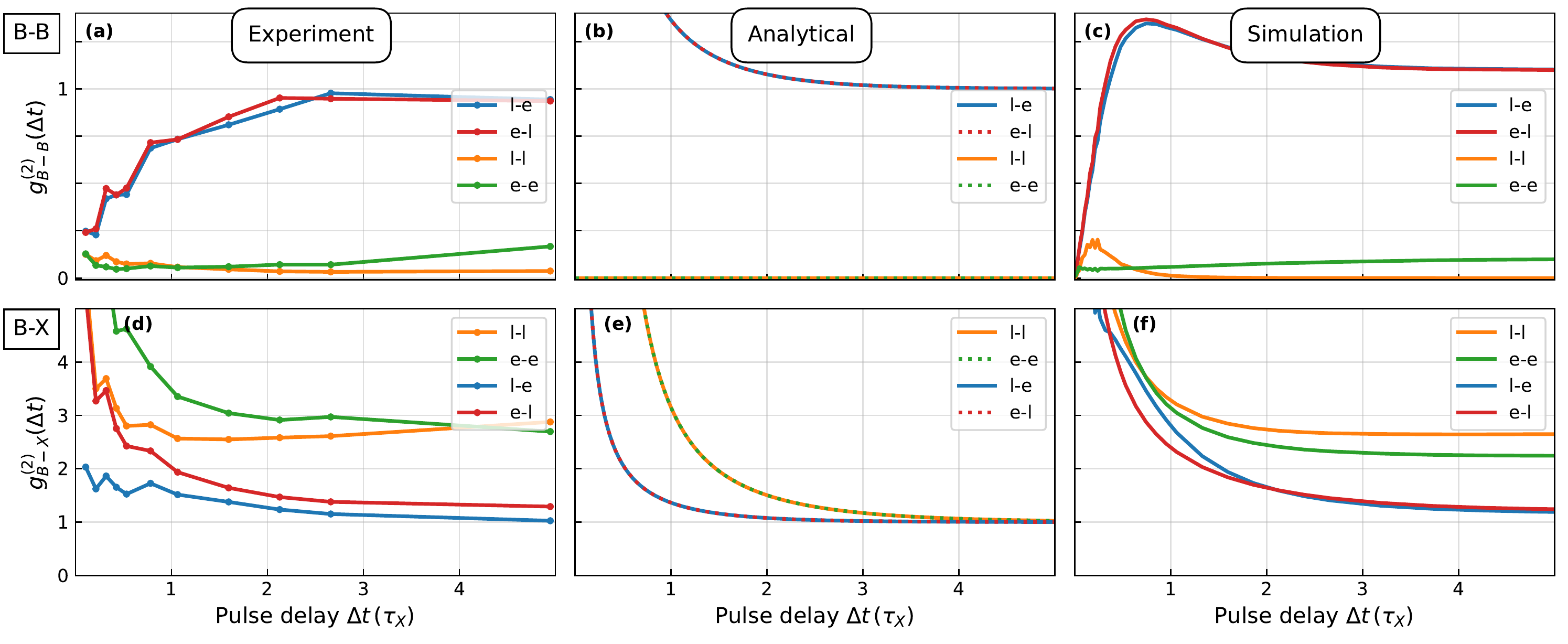}}
    \caption{Normalized time-resolved correlations $g_{i,n-j,m}^{(2)}(\Delta t)$ (a,d) between photons within the same spectral mode ($i=j=B$) in the upper panels, and from different spectral modes ($i=B,j=X$) in the lower panels extracted from our experimental data. For comparison the center (b,e) and right (c,f) panels show the analytical expectation for ideal conditions (from Eq.~\ref{eq:g2}) and the more realistic MC simulations, respectively. The different general trends, observed in our experimental data, are much better reproduced in the MC simulation compared to the analytical expectations.}
\label{fig:extracted_g2}
\end{figure*}

In the following, we provide a detailed analysis of the intensity correlations in each quadrant for the different time- and energy-modes. To this end, we first sum the coincidences in each temporal quadrant, normalised by the corresponding quadrant from an uncorrelated two-time correlation map (arising from detection events separated by 12.5$\,$ns). We emphasize that Figure~\ref{fig:butterflies_data} displays the two-time correlation maps from the same $\ket{\psi}$ wave-packet; the corresponding maps from two uncorrelated $\ket{\psi}$ wave-packets are also measured and used for normalization.
\\
In a next step, the correlation $g_{\text{i,n--j,m}}^{(2)}(\Delta t)$ between different spectral and temporal modes is extracted, both for the experimental data and the MC simulated coincidences. Their dependence on $\Delta t$ is studied in Figure~\ref{fig:extracted_g2}. Here, the first column (panels (a,d)) shows the extracted correlations from the measured two-time correlation maps, the second column (panels (b,e)) displays the analytical calculations based on Eq.~\ref{eq:g2}, and the third column (panels (c,f)) depicts the results of extracting the normalized correlations from the simulated two-time correlation maps, generated by the realistic MC simulation. We find that the experimental results significantly differ from the analytical expectation, mainly due the non-ideal preparation fidelity and the polarization filtering. But when incorporating these experimental imperfections into the MC simulation, the different trends of the experimental data are well reproduced.
We attribute remaining deviations between experiments and MC simulation to non-strictly mono-exponential decays and additional decay channels \cite{lehner2023beyond}, not included in the MC simulation, as well as the influence of detector dark counts (here $\approx 50 \,$Hz), which become increasingly important at small pulse separations associated with low count rates. Additionally, the correlations extracted from the two-time correlation maps also sensitively depend on the exact choice of the origin of the four quadrants. To define the quadrant limits in Figure \ref{fig:butterflies_data} we fit the arrival time of the first pulse and add $\Delta t$ to obtain the time-bin limits, as explained in the Methods section. Note, however, that this choice will be up to the users of a quantum information protocol and can thus be optimized to maximize, e.g., the fidelity to a desired state as done in Ref. \cite{wein2022photon}.
\\

\subsection{Confirming the Purity of the Entangled State}
\label{Sec:HOM-predictions}
The intensity correlations of different spectral and temporal modes are not sufficient to draw conclusions about the purity of the state $\ket{\psi}$. A completely mixed version of the state would provide the same results reported so far (lifetime- and intensity-correlations). While the observation of Rabi rotations during the TPE excitation indicates the initial coherent preparation of a pure biexciton state, dephasing mechanisms might degrade the coherence until the second TPE pulse arrives, resulting in a partially mixed state in the photon-number basis.
\\
Thus, to probe the purity of $\ket{\psi}$ in a high-dimensional space, one can implement the HOM interference of two copies of $\ket{\psi}$, controlling their relative phase $\phi$, as indicated before. This is experimentally challenging for the following reasons: (1) As the energy-entangled state has to be maintained until the interference takes place, spectral filtering of the X- and B-photons is not allowed. (2) The HOM bunching only occurs for photons interfering in the same spectral and temporal mode, which further reduces the visibility, already intrinsically limited by the radiative cascade. (3) The QDs used in our experiment are not embedded in a nano-photonic resonator, which significantly limits the photon collection efficiency, rendering phase-resolved HOM experiments difficult.

Concerning point (1), we note that the spectral filtering of $\ket{\psi}$ from the B- and X-modes would correspond to a partial trace of its density matrix, resulting in individually mixed states, which is why collecting only the X photon under TPE, without a stimulation pulse, leads to the loss of purity in the Fock basis \cite{karli2024controlling}. A preliminary solution is attempted in this work by using a filtering stage that separates the TPE laser from both B and X emission (collected together), before sending the full state to our HOM experiment realized by a path-unbalanced MZI described in the Methods section. The successful filtering of the full state $\ket{\psi}$ is confirmed by measuring the joint HOM interference of the B-X-state resulting from a single TPE excitation.

\begin{figure}[ht]
    \centering
\subfloat{\includegraphics[width=1\linewidth]{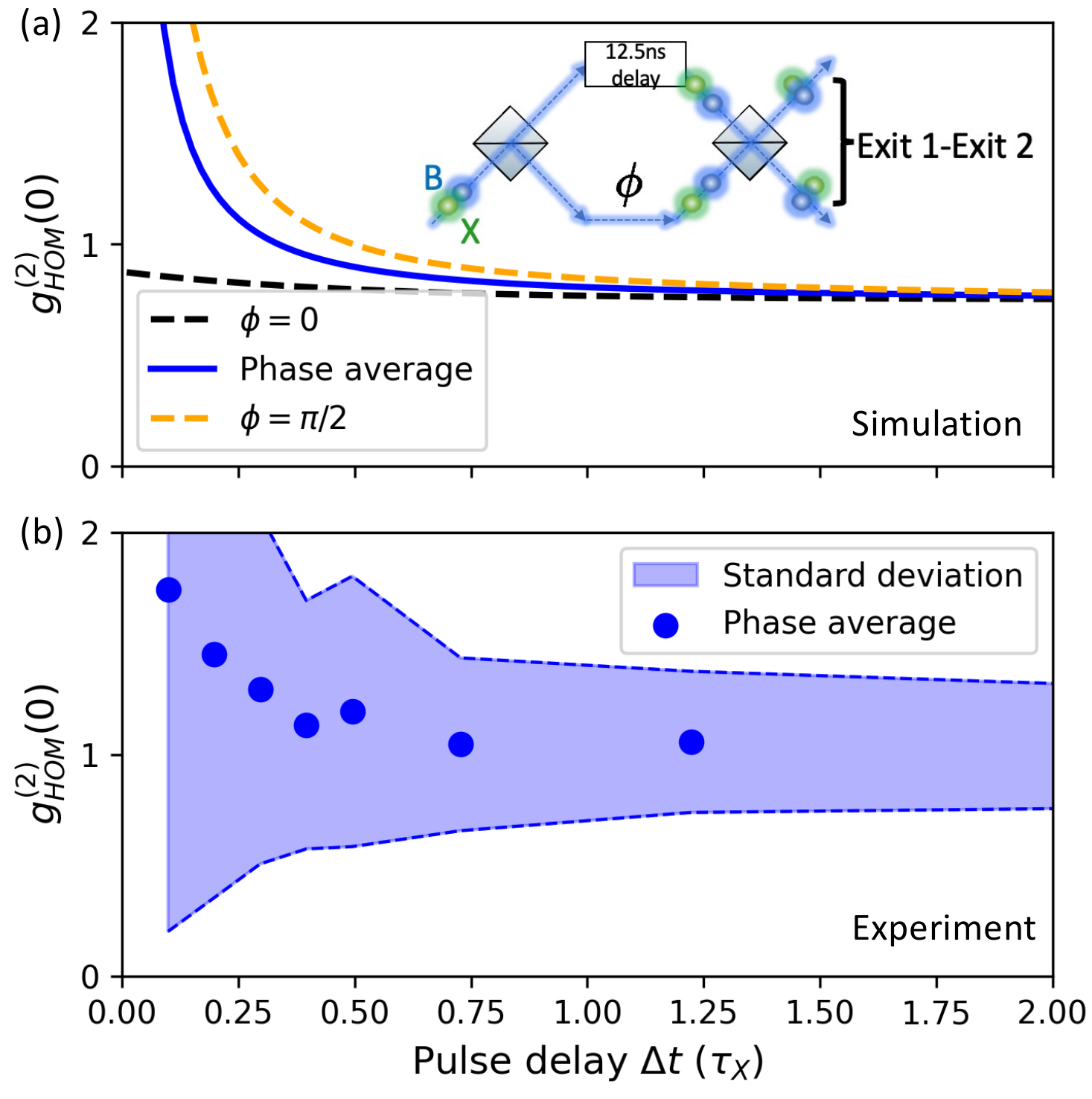}}
    \caption{Phase-controlled HOM interference of the full state $\ket{\psi}$ with its time-delayed copy to partially confirm the purity: (a) Expected HOM correlations obtained from analytical calculations for a fixed relative phase of $0$ (black lines) and $\pi/2$ (orange lines) for the case of resolving only the spatial degrees of freedom (summing over time and energy modes). Inset shows how the correlation can be obtained experimentally by launching the full state (including X- and B-modes) into a MZI with controllable relative phase $\phi$ and measuring coincidences between selected modes at the output. (b) Experimental results for the prediction of (a), obtained without phase-resolution but by averaging over many realizations of different, temporarily fixed phases.}
    \label{fig:HOM}
\end{figure}

The possibility to fully resolve the HOM interference of two copies of $\ket{\psi}$ in energy and time as a function of the relative phase $\phi$ provides different correlations in analogy to the different $g^{(2)}_{i,n-j,m}$ combinations. The strongest influence of phase variations is expected for $\ket{\psi}$ states with significant vacuum contribution {($\Delta t \ll \tau_X$ )}. This however, implies collecting a very low average photon number due to the de-excitation of the biexciton (cf. Figure~\ref{fig:lifetimes}(d)). Such a prominent presence of vacuum severely limits the count-rates in our experiment making a conclusive phase-resolved HOM measurement, in which energy and temporal modes are resolved and fully correlated, very challenging. For this reason, we perform a HOM measurement of the full state $\ket{\psi}$, where $\phi$ evolves randomly and we measure only correlations between the two outputs of the BS without resolving the spectral or temporal modes. Additionally, limiting the integration time of each interference experiment to $\approx 1 \,$s assures phase-stability during the measurement, at the cost of relatively low statistics.
\\
To enable a proper interpretation of our experimental results, we first analyze the predicted behavior of the full state interference in Figure~\ref{fig:HOM}(a). Here, the expected correlations are predicted via analytical calculations (cf. Eq. \ref{eq:hom}), assuming an ideal indistinguishability for the sake of simplicity. Note, that we additionally performed numerical open system simulations accounting for a reduced photon-indistinguishability, due to the timing jitter introduced by the radiative cascade, resulting in the same qualitative behavior.
\\
As mentioned above, even when resolving neither the spectral nor the temporal modes, one can observe an influence of $\phi$ as shown in Figure~\ref{fig:HOM}(a), where the correlation between only the two spatial modes of the BS output (see inset) is predicted. This correlation was experimentally measured by averaging over the results from many repeated interference experiments for approximately fixed phase $\phi$ (1$\,$s integration time) yielding Figure~\ref{fig:HOM}(b). During the calculation of $g^{(2)}_{\text{HOM}}$ from the experimental data we employ resampling techniques by repeatedly shifting the 1$\,$s window by a few ms, and averaging over all results, to increase the statistics as described in the Methods Section. The blue dots represent the average of the HOM correlation values measured for many different randomly occurring $\phi$ values, and follow the expected trend by being a mixture between the extreme curves of Figure~\ref{fig:HOM}(a) (blue line). The spread of measured HOM values for different $\phi$ values should decrease when increasing $\Delta t$ as the vacuum component in Eq.~\ref{eq:state} reduces, reducing its susceptibility to phase fluctuations, which is reflected in the reduced standard deviation (blue shaded region). However, at the same time the spread also reduces due to an improved signal-to-noise ratio at higher delays, rendering quantitative conclusions difficult. In addition, there are also deviations between the predicted results and the measured data caused by the polarization filtering in combination with polarization correlations within one decay-channel of the cascade, which are not included; neither in the analytical model nor the open system simulation. Here, future investigations will be needed.

\section{Discussion}
In this work, we demonstrated a sequential two-photon resonant excitation process for coherently driving a solid-state 3-level system represented by the B-X radiative cascade of a single QD. By performing energy- and time-resolved correlation experiments, we confirm that the resulting state does consist of zero to four photons distributed over two temporal and two spectral modes, and that its composition changes when varying the delay $\Delta t$ between the excitation pulses. The experimentally observed dynamics are well reproduced using extensive modeling, both analytically and numerically, including most experimental imperfections as well as realistic QD parameters in a MC simulation. Finally, interference experiments are proposed to confirm the purity of the created state and a very first experimental realization is presented. Important next steps concern the full tomography of the generated state, which requires improvements on the setup's phase stability as well as an enhanced photon collection efficiency from the quantum emitter. We anticipate that these improvements will pave the way for a future exploitation of entanglement in the photon-number basis for the implementation of advanced schemes in quantum networking as discussed in the following. 
\\
Concerning the practical applicability of the sequential excitation scheme, one can envision different possible applications, to make use of the complex entangled state. On the one hand, making suitable choices of the delay $\Delta t$, the state $\ket{\psi}$ can be split between two parties so that their shared mutual information is higher than that of a GHZ state of the same dimension \cite{santos2023multipartite}. Additionally, for $\Delta t \ll \tau_X$ the state $\ket{\psi}$ mainly consists of the $\ket{0000}$ and $\ket{1001}$ terms. This represents a frequency-resolved version of the state obtained from sequentially exciting the 2LS \cite{wein2022photon}. This state can be used to violate Bell-type inequalities. 
\\
On the other hand, the generated state $\ket{\psi}$ can be useful to quantify loss or gain of quantum channels. It could be used to quantify how much an experimental teleportation protocol deviates from its ideal version, as the discrete variable state produced here approximates the optimal continuous-variable state to test quantum teleportation \cite{sharma2022optimal}. In general, moving from bipartite Bell states to multi-partite entanglement offers a variety of applications \cite{kempe1999multiparticle}, just as increasing the dimension of each local quantum system adds possible use cases \cite{cozzolino2019high,erhard2020advances}. Therefore, by adding additional excitation pulses to the excitation sequence presented in this work, the dimension of each subspace would be $>2$, yielding a multi-partite, high-dimensional entangled-states from the Terra Incognita region in Ref.~\cite{cervera2022experimental}, representing an alternative to the qutrit GHZ-state \cite{erhard2018experimental}.

At present, the experimentally most challenging part towards entanglement distribution using $\ket{\psi}$ is the separation of the different temporal modes. Possible solutions could be fast on-chip switching or passive demultiplexing in combination with post-selection techniques. Exploiting the coherent superposition of single-photon- and vacuum-states is of increasing interest both, for applications \cite{polacchi2023quantum,erkilicc2023surpassing} as well as for the characterization of quantum emitters \cite{neuwirth2022multipair}. In this context, our work represents a significant step towards increasing the complexity and thus power of multi-partite entangled states encoded in the photon-number-basis using deterministic quantum light sources.

\begin{acknowledgments}
The authors acknowledge helpful discussions with Mark Wilde, Thomas Bracht, and Doris Reiter, as well as technical support by Anton Klimek. D.A.V., N.D.K., M.v.H., and T.H. acknowledge financial support by the German Federal Ministry of Education and Research (BMBF) via the project “QuSecure” (Grant~No.~13N14876) within the funding program Photonic Research Germany, the BMBF joint project “tubLAN Q.0” (Grant~No.~16KISQ087K) and the Einstein Foundation via the Einstein Research Unit “Quantum Devices”. C.A.-S. acknowledges the support from the Comunidad de Madrid fund “Atracci\'on de Talento, Mod. 1”, Ref. 2020-T1/IND-19785, the project from the Ministerio de Ciencia e Innovaci\'on PID2020-113445GB-I00, the project ULTRABRIGHT from the Fundaci\'on Ram\'on Areces and the Grant “Leonardo for researchers in Physics 2023” from Fundaci\'on BBVA. Y.K., F.K., V.R., and G.W. acknowledge financial support via the Austrian Science Fund FWF projects W1259 (DK-ALM Atoms, Light, and Molecules), FG 5, TAI-556N (DarkEneT), and I4380 (AEQuDot). A.R. and S.F.C.d.S. acknowledge the FWF projects FG 5, P 30459, I 4320, the Linz Institute of Technology (LIT) and the European Union's Horizon 2020 research and innovation program under Grant Agreement Nos. 899814 (Qurope), 871130 (ASCENT+) and the QauntERA II Programme (project QD-E-QKD).
\end{acknowledgments}

\section*{Methods}
\textbf{Sample:} Our experiments are performed on a single GaAs/AlGaAs QD grown via molecular beam epitaxy employing the local droplet etching method \cite{da2021gaas}. \\\textbf{Setup:} In order to perform the sequential TPE a spectrally tunable ps-laser (picoEmerald, APE) was spectrally sliced in a homemade folded 4f-pulse-shaper to a pulse duration of 6$\,$ps and tuned to the TPE resonance. The QD was kept at a temperature of 4.3$\,K$ in a closed-cycle cryostat (attoDry800 by Attocube Systems) and the emission was collected with an aspheric lens ($NA=0.8$) and coupled into single mode fibers. Spectral selection of single transitions was achieved via notch filters of 0.3$\,$nm full-width half-maximum (FWHM) and the single photons were detected on a superconducting nanowire single photon detector (by Single Quantum) with a timing jitter of $\approx$ 40$\,$ps and detection efficiencies around $80\%$, before the time-tags were digitally saved by a time-to-digital-converter (quTag by Qutools). The second excitation pulse was introduced by an asymmetric Michelson-interferometer with a variable relative delay $\Delta t$ and equal transmission of both arms. A motorized delay (LTS150C by Thorlabs) allows for adjusting maximum pulse delays of 2$\,$ns with a precision of 3$\,$ps and its linear behavior was confirmed in calibration measurements.
\\ \\
\textbf{Two-time Correlation Measurements:} The correlation measurements were carried out by spectrally filtering both X- and B-photons, sending each to an balanced 1:2 fiber-coupler for detecting the photons on four parallel detector channels with $\approx 80\%$ detection efficiency. The full timestamps are saved with an integration time of 10$\,$min for each delay $\Delta t$, allowing to retrieve all possible combinations of correlations, as well as the time dynamics from the same dataset afterwards. To set the limits of the four quadrants describing the different temporal mode combinations, we fit the arrival time distributions with a product of a Heaviside-function and an exponential decay which is convoluted with the system response function to determine the exact arrival time of the first $\pi$-pulse. Adding the pulse separation $\Delta t$ to this value, we obtain the time bin separation used to define the quadrants. The normalized correlations are obtained by dividing the integral of a quadrant in the two-time correlation map by the corresponding integral for a correlation map between uncorrelated photons separated by 12.5$\,$ns. 
\\ \\
\textbf{HOM Measurements} The HOM experiments are performed in a path-unbalanced MZI without active phase-stabilization. A delay of 12.5$\,$ns (matching the repetition rate of the laser) in one arm allows to prepare two copies of the entangled state to interfere in a fiber BS. Fiber polarization controllers ensure the co-polarized interference. Interference of the full state $\ket{\psi}$ is achieved by filtering the QD emission (including both the B- and X-photons) with a bandpass filter of 3$\,$nm FWHM in combination with a set of notch filters to remove residual scattered laser light in between the B- and X-lines. The HOM interferometer showed almost ideal interference visibility ($\approx 100\%$) using attenuated laser pulses at the same wavelength, confirming the precise timing and polarization control in our setup. 
\\ \\
\textbf{Monte Carlo Simulations:} Stochastic Monte Carlo simulation were employed to incorporate experimental imperfections into the two-time correlation maps. This allows us to explore the effect of different experimental imperfections. The simulations provided in the main text used the following parameters which were determined experimentally in separate characterization measurements: Preparation fidelity $\mathcal{F}_{\text{prep}} = 0.9 $, detector timing jitter $\delta t = 40\,$ps, detector deadtime $t_{Dt} = 100\,$ns, the laser pulse duration $\tau_{\text{laser}}=6\,$ps. 
\clearpage
\newpage
\appendix


\begin{thebibliography}{10}

\bibitem{santos2023multipartite}
A.~C. Santos, C.~Schneider, R.~Bachelard, A.~Predojevi{\'c}, and
  C.~Ant{\'o}n-Solanas.
\newblock Multipartite entanglement encoded in the photon-number basis by
  sequential excitation of a three-level system.
\newblock {\em Optics Letters}, 48(23):6332--6335, 2023.

\bibitem{kimble2008quantum}
H.~J. Kimble.
\newblock The quantum internet.
\newblock {\em Nature}, 453(7198):1023--1030, 2008.

\bibitem{thomas2022efficient}
P.~Thomas, L.~Ruscio, O.~Morin, and G.~Rempe.
\newblock Efficient generation of entangled multiphoton graph states from a
  single atom.
\newblock {\em Nature}, 608(7924):677--681, 2022.

\bibitem{akopian2006entangled}
N.~Akopian, N.~H. Lindner, E.~Poem, Y.~Berlatzky, J.~Avron, D.~Gershoni, B.~D.
  Gerardot, and P.~M. Petroff.
\newblock Entangled photon pairs from semiconductor quantum dots.
\newblock {\em Physical review letters}, 96(13):130501, 2006.

\bibitem{huber2017highly}
D.~Huber, M.~Reindl, Y.~Huo, H.~Huang, J.~S. Wildmann, O.~G. Schmidt,
  A.~Rastelli, and R.~Trotta.
\newblock Highly indistinguishable and strongly entangled photons from
  symmetric gaas quantum dots.
\newblock {\em Nature communications}, 8(1):15506, 2017.

\bibitem{basset2019entanglement}
F.~Basso~Basset, M.~B. Rota, C.~Schimpf, D.~Tedeschi, K.~D. Zeuner, S.~F.~C.
  Da~Silva, M.~Reindl, V.~Zwiller, K.~D. J{\"o}ns, A.~Rastelli, and R.~Trotta.
\newblock Entanglement swapping with photons generated on demand by a quantum
  dot.
\newblock {\em Physical Review Letters}, 123(16):160501, 2019.

\bibitem{basso2021quantum}
F.~Basso~Basset, M.~Valeri, E.~Roccia, V.~Muredda, D.~Poderini, J.~Neuwirth,
  N.~Spagnolo, M.~B. Rota, G.~Carvacho, F.~Sciarrino, and R.~Trotta.
\newblock Quantum key distribution with entangled photons generated on demand
  by a quantum dot.
\newblock {\em Science advances}, 7(12):eabe6379, 2021.

\bibitem{vajner2022quantum}
D.~A. Vajner, L.~Rickert, T.~Gao, K.~Kaymazlar, and T.~Heindel.
\newblock Quantum communication using semiconductor quantum dots.
\newblock {\em Advanced Quantum Technologies}, 5(7):2100116, 2022.

\bibitem{muller2009creating}
A.~Muller, W.~Fang, J.~Lawall, and G.~S Solomon.
\newblock Creating polarization-entangled photon pairs from a semiconductor
  quantum dot using the optical stark effect.
\newblock {\em Physical review letters}, 103(21):217402, 2009.

\bibitem{bennett2010electric}
A.~J. Bennett, y~M.~A. Pooley, n~R.~M. Stevenson, M.~B. Ward, R.~Patel, Boyer,
  A.~de~La~Giroday, N.~Sk{\"o}ld, I.~Farrer, C.~A. Nicoll, D.~A. Ritchie, and
  A.~J. Shields.
\newblock Electric-field-induced coherent coupling of the exciton states in a
  single quantum dot.
\newblock {\em Nature Physics}, 6(12):947--950, 2010.

\bibitem{huber2018strain}
D.~Huber, M.~Reindl, S.~F. Covre~da Silva, C.~Schimpf,
  J.~Mart{\'\i}n-S{\'a}nchez, H.~Huang, G.~Piredda, Johannes Edlinger,
  A.~Rastelli, and R.~Trotta.
\newblock Strain-tunable gaas quantum dot: A nearly dephasing-free source of
  entangled photon pairs on demand.
\newblock {\em Physical review letters}, 121(3):033902, 2018.

\bibitem{seidelmann2022two}
T.~Seidelmann, C.~Schimpf, T.~K. Bracht, M.~Cosacchi, A.~Vagov, A.~Rastelli,
  D.~E. Reiter, and V.~M. Axt.
\newblock Two-photon excitation sets limit to entangled photon pair generation
  from quantum emitters.
\newblock {\em Physical Review Letters}, 129(19):193604, 2022.

\bibitem{scholl2020crux}
E.~Sch{\"o}ll, L.~Schweickert, L.~Hanschke, K.~D. Zeuner, F.~Sbresny,
  T.~Lettner, R.~Trivedi, M.~Reindl, S.~F. Covre Da~Silva, R.~Trotta, J.~F.
  Finley, J~Vučković, K~Müller, A.~Rastelli, Val Zwiller, and K.~D. Jöns.
\newblock Crux of using the cascaded emission of a three-level quantum ladder
  system to generate indistinguishable photons.
\newblock {\em Physical Review Letters}, 125(23):233605, 2020.

\bibitem{jayakumar2014time}
H.~Jayakumar, A.~Predojevi{\'c}, T.~Kauten, T.~Huber, G.~S. Solomon, and
  G.~Weihs.
\newblock Time-bin entangled photons from a quantum dot.
\newblock {\em Nature communications}, 5(1):4251, 2014.

\bibitem{huber2016coherence}
T.~Huber, L.~Ostermann, M.~Prilm{\"u}ller, G.~S. Solomon, H.~Ritsch, G.~Weihs,
  and A.~Predojevi{\'c}.
\newblock Coherence and degree of time-bin entanglement from quantum dots.
\newblock {\em Physical Review B}, 93(20):201301, 2016.

\bibitem{gines2021time}
L.~Gin{\'e}s, C.~Pepe, J.~Gonzales, N.~Gregersen, S.~H{\"o}fling, C.~Schneider,
  and A.~Predojevi{\'c}.
\newblock Time-bin entangled photon pairs from quantum dots embedded in a
  self-aligned cavity.
\newblock {\em Optics Express}, 29(3):4174--4180, 2021.

\bibitem{prilmuller2018hyperentanglement}
M.~Prilm{\"u}ller, T.~Huber, M.~M{\"u}ller, P.~Michler, G.~Weihs, and
  A.~Predojevi{\'c}.
\newblock Hyperentanglement of photons emitted by a quantum dot.
\newblock {\em Physical Review Letters}, 121(11):110503, 2018.

\bibitem{schwartz2016deterministic}
I.~Schwartz, D.~Cogan, E.~R. Schmidgall, Y.~Don, L.~Gantz, O.~Kenneth, N.~H.
  Lindner, and D.~Gershoni.
\newblock Deterministic generation of a cluster state of entangled photons.
\newblock {\em Science}, 354(6311):434--437, 2016.

\bibitem{cogan2023deterministic}
D.~Cogan, Z.~Su, O.~Kenneth, and D.~Gershoni.
\newblock Deterministic generation of indistinguishable photons in a cluster
  state.
\newblock {\em Nature Photonics}, 17(4):324--329, 2023.

\bibitem{tiurev2022high}
Konstantin Tiurev, Martin~Hayhurst Appel, Pol~Llopart Mirambell, Mikkel~Bloch
  Lauritzen, Alexey Tiranov, Peter Lodahl, and Anders~S{\o}ndberg S{\o}rensen.
\newblock High-fidelity multiphoton-entangled cluster state with solid-state
  quantum emitters in photonic nanostructures.
\newblock {\em Physical Review A}, 105(3):L030601, 2022.

\bibitem{coste2023high}
N.~Coste, D.~A. Fioretto, N.~Belabas, S.~C. Wein, P.~Hilaire, R.~Frantzeskakis,
  M.~Gundin, B.~Goes, N.~Somaschi, M.~Morassi, A.~Lemaitre, I.~Sagnes,
  A.~Harouri, S.~E. Economou, A.~Auffeves, O.~Krebs, L.~Lanco, and
  P.~Senellart.
\newblock High-rate entanglement between a semiconductor spin and
  indistinguishable photons.
\newblock {\em Nature Photonics}, pages 1--6, 2023.

\bibitem{browne2005resource}
D.~E. Browne and T.~Rudolph.
\newblock Resource-efficient linear optical quantum computation.
\newblock {\em Physical Review Letters}, 95(1):010501, 2005.

\bibitem{luker2015direct}
S.~L{\"u}ker, T.~Kuhn, and D.~E. Reiter.
\newblock Direct optical state preparation of the dark exciton in a quantum
  dot.
\newblock {\em Physical Review B}, 92(20):201305, 2015.

\bibitem{kappe2024keeping}
Florian Kappe, Ren{\'e} Schwarz, Yusuf Karli, Thomas Bracht, Vollrath~M Axt,
  Armando Rastelli, Vikas Remesh, Doris~E Reiter, and Gregor Weihs.
\newblock Keeping the photon in the dark: Enabling full quantum dot control by
  chirped pulses and magnetic fields.
\newblock {\em arXiv preprint arXiv:2404.10708}, 2024.

\bibitem{istrati2020sequential}
D.~Istrati, Y.~Pilnyak, J.~C. Loredo, C.~Ant{\'o}n-Solanas, N.~Somaschi,
  P.~Hilaire, H.~Ollivier, M.~Esmann, L.~Cohen, L.~Vidro, C.~Millet,
  A.~Lemaitre, I.~Sagnes, A.~Harouri, L.~Lanco, P.~Senellart, and H.~S.
  Eisenberg.
\newblock Sequential generation of linear cluster states from a single photon
  emitter.
\newblock {\em Nature communications}, 11(1):5501, 2020.

\bibitem{maring2024versatile}
N.~Maring, A.~Fyrillas, M.~Pont, E.~Ivanov, P.~Stepanov, N.~Margaria, W.~Hease,
  A.~Pishchagin, A.~Lema{\^\i}tre, I.~Sagnes, T.~H. Au, S.~Boissier,
  E.~Bertasi, A.~Baert, M.~Baldivia, M.~Billard, O.~Acar, A.~Brieussel,
  R.~Mezher, S.~C. Wein, A.~Salavrakos, P.~Sinnott, D.~A. Fioretto, Emeriaum
  P.-E., N.~Belabas, S.~Mansfield, P.~Senellart, J.~Senellart, and N.~Somaschi.
\newblock A versatile single-photon-based quantum computing platform.
\newblock {\em Nature Photonics}, pages 1--7, 2024.

\bibitem{cao2024photonic}
H.~Cao, L.~M. Hansen, F.~Giorgino, L.~Carosini, P.~Zah\'alka, F.~Zilk, J.~C.
  Loredo, and P.~Walther.
\newblock Photonic source of heralded greenberger-horne-zeilinger states.
\newblock {\em Phys. Rev. Lett.}, 132:130604, Mar 2024.

\bibitem{arrazola2014quantum}
Juan~Miguel Arrazola and Norbert L{\"u}tkenhaus.
\newblock Quantum fingerprinting with coherent states and a constant mean
  number of photons.
\newblock {\em Physical Review A}, 89(6):062305, 2014.

\bibitem{de2023experimental}
I~Maillette de~Buy~Wenniger, SE~Thomas, M~Maffei, SC~Wein, M~Pont, N~Belabas,
  S~Prasad, A~Harouri, A~Lema{\^\i}tre, I~Sagnes, N.~Somaschi, A.~Auffeves, and
  P.~Senellart.
\newblock Experimental analysis of energy transfers between a quantum emitter
  and light fields.
\newblock {\em Physical Review Letters}, 131(26):260401, 2023.

\bibitem{munoz2014emitters}
C~S{\'a}nchez Mu{\~n}oz, E~Del~Valle, A~Gonz{\'a}lez Tudela, K~M{\"u}ller,
  S~Lichtmannecker, M~Kaniber, C~Tejedor, JJ~Finley, and FP~Laussy.
\newblock Emitters of n-photon bundles.
\newblock {\em Nature photonics}, 8(7):550--555, 2014.

\bibitem{munoz2018filtering}
Carlos~S{\'a}nchez Mu{\~n}oz, Fabrice~P Laussy, Elena del Valle, Carlos
  Tejedor, and Alejandro Gonz{\'a}lez-Tudela.
\newblock Filtering multiphoton emission from state-of-the-art cavity quantum
  electrodynamics.
\newblock {\em Optica}, 5(1):14--26, 2018.

\bibitem{renema2020simulability}
Jelmer~J Renema.
\newblock Simulability of partially distinguishable superposition and gaussian
  boson sampling.
\newblock {\em Physical Review A}, 101(6):063840, 2020.

\bibitem{loredo2019generation}
J.~C. Loredo, C.~Ant{\'o}-Solanas, B.~Reznychenko, P.~Hilaire, A.~Harouri,
  C.~Millet, H.~Ollivier, N.~Somaschi, L.~De~Santis, A~Lema{\^\i}tre,
  I.~Sagnes, L.~Lanco, A.~Auffeves, O.~Krebs, and P.~Senellart.
\newblock Generation of non-classical light in a photon-number superposition.
\newblock {\em Nature Photonics}, 13(11):803--808, 2019.

\bibitem{bozzio2022enhancing}
M.~Bozzio, M.~Vyvlecka, M.~Cosacchi, C.~Nawrath, T.~Seidelmann, J.~C. Loredo,
  S.~L. Portalupi, V.~M. Axt, P.~Michler, and P.~Walther.
\newblock Enhancing quantum cryptography with quantum dot single-photon
  sources.
\newblock {\em npj Quantum Information}, 8(1):104, 2022.

\bibitem{karli2024controlling}
Y.~Karli, D.~A. Vajner, F.~Kappe, P.~C.~A. Hagen, L.~M. Hansen, R.~Schwarz,
  T.~K. Bracht, C.~Schimpf, S.~F. Covre~da Silva, P.~Walther, A.~Rastelli,
  V.~M. Axt, J.~C. Loredo, V.~Remesh, T.~Heindel, D.~Reiter, and G.~Weihs.
\newblock Controlling the photon number coherence of solid-state quantum light
  sources for quantum cryptography.
\newblock {\em npj Quantum Information}, 10(1):17, 2024.

\bibitem{erkilicc2023surpassing}
O.~Erk{\i}l{\i}{\c{c}}, L.~Conlon, B.~Shajilal, S.~Kish, S.~Tserkis, Y.-S. Kim,
  P.~K. Lam, and S.~M. Assad.
\newblock Surpassing the repeaterless bound with a photon-number encoded
  measurement-device-independent quantum key distribution protocol.
\newblock {\em npj Quantum Information}, 9(1):29, 2023.

\bibitem{polacchi2023quantum}
B.~Polacchi, F.~Hoch, G.~Rodari, S.~Savo, G.~Carvacho, N.~Spagnolo,
  T.~Giordani, and F.~Sciarrino.
\newblock Quantum teleportation of a genuine vacuum-one-photon qubit generated
  via a quantum dot source.
\newblock {\em arXiv preprint arXiv:2310.20521}, 2023.

\bibitem{wenniger2024photonic}
I.~Wenniger, S.~C. Wein, D.~Fioretto, S.~E. Thomas, C.~Ant{\'o}n-Solanas,
  A.~Lema{\^\i}tre, I.~Sagnes, A.~Harouri, N.~Belabas, N.~Somaschi, P.~Hilaire,
  J.~Senellart, and P.~Senellart.
\newblock Photonic quantum interference in the presence of coherence with
  vacuum.
\newblock {\em arXiv preprint arXiv:2401.01187}, 2024.

\bibitem{wein2022photon}
S.~C. Wein, J.~C. Loredo, M.~Maffei, P.~Hilaire, A.~Harouri, N.~Somaschi,
  A.~Lema{\^\i}tre, I.~Sagnes, L.~Lanco, O.~Krebs, A~Auffeves, C.~Simon,
  P~Senellart, and C~Ant{\'o}n-Solanas.
\newblock Photon-number entanglement generated by sequential excitation of a
  two-level atom.
\newblock {\em Nature Photonics}, 16(5):374--379, 2022.

\bibitem{jayakumar2013deterministic}
H.~Jayakumar, A.~Predojevi{\'c}, T.~Huber, T.~Kauten, G.~S. Solomon, and
  G.~Weihs.
\newblock Deterministic photon pairs and coherent optical control of a single
  quantum dot.
\newblock {\em Physical review letters}, 110(13):135505, 2013.

\bibitem{muller2014demand}
M.~M{\"u}ller, S.~Bounouar, K.~D. J{\"o}ns, M.~Gl{\"a}ssl, and P.~Michler.
\newblock On-demand generation of indistinguishable polarization-entangled
  photon pairs.
\newblock {\em Nature Photonics}, 8(3):224--228, 2014.

\bibitem{bauch2023demand}
D.~Bauch, D.~Siebert, K.~D. J{\"o}ns, J.~F{\"o}rstner, and S.~Schumacher.
\newblock On-demand indistinguishable and entangled photons using tailored
  cavity designs.
\newblock {\em Advanced Quantum Technologies}, page 2300142, 2023.

\bibitem{reischle2008influence}
M~Reischle, GJ~Beirne, R~Ro{\ss}bach, M~Jetter, and P~Michler.
\newblock Influence of the dark exciton state on the optical and quantum
  optical properties of single quantum dots.
\newblock {\em Physical review letters}, 101(14):146402, 2008.

\bibitem{wang2019demand}
H.~Wang, H.~Hu, T.-H. Chung, J.~Qin, X.~Yang, J.-P. Li, R.-Z. Liu, H.-S. Zhong,
  Y.-M. He, X.~Ding, Y.-H. Deng, Q.~Dai, Y.-H. Hui, S.~Höfling, C.-Y. Lu, and
  J.-W. Pan.
\newblock On-demand semiconductor source of entangled photons which
  simultaneously has high fidelity, efficiency, and indistinguishability.
\newblock {\em Physical review letters}, 122(11):113602, 2019.

\bibitem{neuwirth2022multipair}
J.~Neuwirth, F.~Basso~Basset, M.~B. Rota, J.-G. Hartel, M.~Sartison, S.~F.
  Covre~da Silva, K.~D. J{\"o}ns, A.~Rastelli, and R.~Trotta.
\newblock Multipair-free source of entangled photons in the solid state.
\newblock {\em Physical Review B}, 106(24):L241402, 2022.

\bibitem{vajner2024demand}
D.~A. Vajner, P.~Holewa, E.~Zieba-Ostoj, M.~Wasiluk, M.~von Helversen,
  A.~Sakanas, A.~Huck, K.~Yvind, N.~Gregersen, A.~Musia{\l}, M.~Syperek,
  E.~Semenova, and T.~Heindel.
\newblock On-demand generation of indistinguishable photons in the telecom
  c-band using quantum dot devices.
\newblock {\em ACS photonics}, 2024.

\bibitem{liu2019solid}
J.~Liu, R.~Su, Y.~Wei, B.~Yao, S.~F. Covre~da Silva, Y.~Yu, J.~Iles-Smith,
  K.~Srinivasan, A.~Rastelli, J.~Li, and X.~Wang.
\newblock A solid-state source of strongly entangled photon pairs with high
  brightness and indistinguishability.
\newblock {\em Nature nanotechnology}, 14(6):586--593, 2019.

\bibitem{brown1956correlation}
R.~Hanbury~Brown and R.~Q. Twiss.
\newblock Correlation between photons in two coherent beams of light.
\newblock {\em Nature}, 177(4497):27--29, 1956.

\bibitem{lehner2023beyond}
Barbara~Ursula Lehner, Tim Seidelmann, Gabriel Undeutsch, Christian Schimpf,
  Santanu Manna, Micha{\l} Gawe{\l}czyk, Saimon~Filipe Covre~da Silva, Xueyong
  Yuan, Sandra Stroj, Doris~E Reiter, et~al.
\newblock Beyond the four-level model: dark and hot states in quantum dots
  degrade photonic entanglement.
\newblock {\em Nano Letters}, 23(4):1409--1415, 2023.

\bibitem{sharma2022optimal}
Kunal Sharma, Barry~C Sanders, and Mark~M Wilde.
\newblock Optimal tests for continuous-variable quantum teleportation and
  photodetectors.
\newblock {\em Physical Review Research}, 4(2):023066, 2022.

\bibitem{kempe1999multiparticle}
Julia Kempe.
\newblock Multiparticle entanglement and its applications to cryptography.
\newblock {\em Physical Review A}, 60(2):910, 1999.

\bibitem{cozzolino2019high}
Daniele Cozzolino, Beatrice Da~Lio, Davide Bacco, and Leif~Katsuo Oxenl{\o}we.
\newblock High-dimensional quantum communication: benefits, progress, and
  future challenges.
\newblock {\em Advanced Quantum Technologies}, 2(12):1900038, 2019.

\bibitem{erhard2020advances}
Manuel Erhard, Mario Krenn, and Anton Zeilinger.
\newblock Advances in high-dimensional quantum entanglement.
\newblock {\em Nature Reviews Physics}, 2(7):365--381, 2020.

\bibitem{cervera2022experimental}
Alba Cervera-Lierta, Mario Krenn, Al{\'a}n Aspuru-Guzik, and Alexey Galda.
\newblock Experimental high-dimensional {Greenberger-Horne-Zeilinger}
  entanglement with superconducting transmon qutrits.
\newblock {\em Physical Review Applied}, 17(2):024062, 2022.

\bibitem{erhard2018experimental}
Manuel Erhard, Mehul Malik, Mario Krenn, and Anton Zeilinger.
\newblock Experimental greenberger--horne--zeilinger entanglement beyond
  qubits.
\newblock {\em Nature Photonics}, 12(12):759--764, 2018.

\bibitem{da2021gaas}
S.~F. Covre~da Silva, G.~Undeutsch, B.~Lehner, S.~Manna, T.~M. Krieger,
  M.~Reindl, C.~Schimpf, R.~Trotta, and A.~Rastelli.
\newblock Gaas quantum dots grown by droplet etching epitaxy as quantum light
  sources.
\newblock {\em Applied Physics Letters}, 119(12), 2021.

\end{thebibliography}
\end{document}